# Graph-based Faulted Line Identification Using Micro-PMU Data in Distribution Systems

Ying Zhang, *Student Member, IEEE*, Jianhui Wang, *Senior Member, IEEE*, and Mohammad Khodayar, *Senior Member, IEEE*

***Abstract*—Motivated by increasing penetration of distributed generators (DGs) and fast development of micro-phasor measurement units (μPMUs), this paper proposes a novel graph-based faulted line identification algorithm using a limited number of μPMUs in distribution networks. The core of the proposed method is to apply advanced distribution system state estimation (DSSE) techniques integrating μPMU data to the fault location. We propose a distributed DSSE algorithm to efficiently restrict the searching region for the fault source in the feeder between two adjacent μPMUs. Based on the graph model of the feeder in the reduced searching region, we further perform the DSSE in a hierarchical structure and identify the location of the fault source. Also, the proposed approach captures the impact of DGs on distribution system operation and remains robust against high-level noises in measurements. Numerical simulations verify the accuracy and efficiency of the proposed method under various fault scenarios covering multiple fault types and fault impedances.**

***Index Terms*— Faulted line identification, distribution systems, state estimation, phasor measurement units, micro-PMUs, distributed generation, graph theory.**

## I. Introduction

FAULTS are regarded as an important type of reliability events, which may immensely affect normal system operation. In the past decade, 22.2 million customers in California experienced about 6000 outage hours resulted from sustained faults [1]. Extensive studies on fault location are developed in meshed transmission systems (*e.g.*, [2] – [4]). However, distribution systems are largely different from transmission systems due to their radial topology and limited real-time meters. Consequently, these existing fault location methods in transmission systems cannot be applied to distribution systems. On the other hand, increasingly pervasive installation of distribution-level phasor measurement units, *i.e.*, micro-PMUs (μPMUs), improves the system monitoring significantly. Compared with conventional meters, μPMUs provide more accurate measurements of voltage and current phasors at a high resolution. Several emerging applications of μPMUs include distribution system state estimation (DSSE), fault detection, and faulted line location [5]. For instance, the authors of [6] applied data-driven techniques with μPMU data to detect the presence of a fault in distribution systems; however, these detection algorithms cannot identify the location of the faulted line.

Y. Zhang, J. Wang, and M. Khodayar are with the Department of Electrical and Computer Engineering, Southern Methodist University, Dallas, TX, USA 75205 (e-mail: yzhang1@smu.edu; jianhui@smu.edu; mkhodayar@smu.edu).

Quick and accurate location of faults in distribution systems helps the utilities to clear the faults and accelerate the system restoration; however, this is a challenging task as the mal-trip or fail-to-trip of the protection devices may lead to inaccurate location of the fault. The chances of such unfavorable events grow with the bidirectional power flow and the increasing penetration of distributed generators (DGs) [7]. The authors of [8] pointed out that conventional protection devices such as fault indicators may fail to clear a fault under the bidirectional current flow conditions. Also, the overcurrent-based protection devices may not be able to locate high-impedance faults in distribution systems since it is difficult to identify the small fault currents [9].

The existing fault location methods are classified into three main types: 1) traveling wave-based, 2) training-based, and 3) impedance-based. The traveling wave-based algorithms (*e.g.*, [10] – [12]) locate a fault by utilizing the arrival time of the original and reflected waves generated by the fault. These methods require high-speed communication and high sampling rate measurements that may not be prevalent in distribution networks. Besides, the training-based fault location methods, such as artificial neural network (ANN) [13] and support vector machine (SVM) [14], require a large number of high-quality measurements as training datasets and thus suffer from a high computational burden in a training process.

Recent efforts are devoted to proposing impedance-based location methods in distribution systems [8], [15] – [21]. For instance, the authors of [17] and [18] proposed the fault location methods focusing on single-phase to ground faults. However, DGs are not considered in the test feeders of [17], while [18] can only localize a faulted area, rather than yielding an exact faulted line. Emerging works are applying μPMU measurements to fault location by constructing generalized impedance-based location methods [8], [19] – [21]. We conclude that the search strategy in these works is to select each bus or each line as the candidate fault source and then calculate the values of the self-defined objective function for each candidate. Then, the fault location is determined by minimizing or maximizing these function values. Specifically, the authors of [19] used a state estimation technique with sufficient μPMU data to identify the fault at a distribution line. However, these μPMUs are assumed available at each bus, which is impractical due to economic and technical restrictions in distribution systems. Further efforts are put into fault location with a fewer number of μPMUs, such as [20] and [21]. The approach in [21] requires equipping with μPMUs at all DGs. This arrangement may not be practical due to a limited number of available $\mu$PMUs. Also, this method does not consider high-impedance faults, which are regarded as an untraceable fault type in system operation. The authors of [21] defined a fault as a generalized reliability event and presented an optimization model to locate the event bus by $\mu$PMU data and pseudo-measurements recorded at load/DG buses. Also, due to the presence of local minimums in the objective function, the method needs to compare all local minimums to obtain a global one. Further, the global minimum of this function points to the final faulted bus. However, this process may increase computational complexity due to this traversal search strategy.

Various influence factors, such as fault types, fault impedances, DG penetration, and measurement errors, may degrade the effectiveness of the existing fault location methods [21]. To mitigate these impacts, this paper proposes a graph-based fault location method using advanced DSSE techniques with $\mu$PMU data in unbalanced distribution systems. The core of the proposed method is to determine the faulted line by comparing the weighted measurement residuals (WMRs) of DSSE in different topologies/graphs. This idea, as a typical application of state estimation, is proposed in [2], [3], and [19], where the power systems are observable by an adequate number of PMUs. In comparison, the proposed method only requires a limited number of $\mu$PMUs in distribution networks for such an application. Specifically, we present an efficient distributed DSSE algorithm to restrict the search region in a shorter feeder between two adjacent $\mu$PMUs. Further, in the shorter feeder, the fault source is identified at the exact line by applying the DSSE methods to a hierarchical structure. The hierarchical structure built on the graph theory is presented in Section III and captures the graphs, subgraphs, and paths in the network.

We list the contributions of this method below:

- The proposed algorithm uses data from limited $\mu$PMUs for fault location. Compared with [19] and [20], this method does not require installing $\mu$PMUs at each node or all DGs.
- This method locates the fault with only during-fault $\mu$PMU data, compared with [20] and [21] that require both pre-fault and during-fault data. Also, unlike the traversal search strategy with high computational complexity in [19]–[21], this method running in a distributed manner has lower computation cost and enables fast faulted line identification within several tens of milliseconds.
- Our approach considers the impact of DG penetration on distribution system operation, and its location performance is independent of fault types and fault impedances. Furthermore, the proposed algorithm is robust against high-level noises in measurements.

## II. THEORETICAL BASIS

This section describes the theoretical basis for applying DSSE to fault location. We introduce a classical state estimator and further extend it to an advanced DSSE method using measurements from a limited number of $\mu$PMUs.

The classical state estimator formulates the relationship between measurements and state variables by

$$\boldsymbol{z} = \boldsymbol{h}(\boldsymbol{x}) + \boldsymbol{e} \tag{1}$$

where $\boldsymbol{x} \in \mathbb{R}^{n\times 1}$ denotes a state variable vector, and $\boldsymbol{z} \in \mathbb{R}^{m\times 1}$ is a measurement vector; $\boldsymbol{h}(\boldsymbol{x})$ is the measurement function, and the measurement noise vector $\boldsymbol{e}$ usually obeys Gaussian distributions $\boldsymbol{e}\sim N(0,\boldsymbol{R})$, where $\boldsymbol{R}$ denotes a covariance matrix, $\boldsymbol{R} = diag[\sigma_1^2, \sigma_2^2, \dots, \sigma_m^2]$, and $\sigma_j^2$ is the variance of the $j$th measurement noise, $j = 1, 2, \dots, m$.

The weighted least square (WLS) criterion is used to minimize the WMR:

$$J = [\boldsymbol{z} - \boldsymbol{h}(\boldsymbol{x})]^T \boldsymbol{W} [\boldsymbol{z} - \boldsymbol{h}(\boldsymbol{x})] \tag{2}$$

where $\boldsymbol{W}$ represents the weight matrix of these measurements, and $\boldsymbol{W} = \boldsymbol{R}^{-1}$.

The estimated states are obtained iteratively by the Gauss-Newton method until each component of $\Delta \boldsymbol{x}^t$ at iteration $t$ is sufficiently small. The process for updating the states is listed as follows:

$$\partial J / \partial \boldsymbol{x}^t = 0 \tag{3}$$

$$\boldsymbol{H}(\boldsymbol{x}^t)^T \boldsymbol{W} \boldsymbol{H}(\boldsymbol{x}^t) \Delta \boldsymbol{x}^t = \boldsymbol{H}(\boldsymbol{x}^t)^T \boldsymbol{W} [\boldsymbol{z} - \boldsymbol{h}(\boldsymbol{x}^t)] \tag{4}$$

$$\boldsymbol{x}^{t+1} = \boldsymbol{x}^t + \Delta \boldsymbol{x}^t \tag{5}$$

where $\boldsymbol{H}(\boldsymbol{x}^t)$ denotes the Jacobian matrix of the measurement function and is calculated by $\partial \boldsymbol{h}(\boldsymbol{x}^t)/\partial \boldsymbol{x}^t$; the symbol $[\cdot]^T$ denotes the transpose of a matrix.

Developed from this classical estimator, the branch current based DSSE method integrating $\mu$PMU data is regarded as a computationally efficient method due to its constant and sparse-structured Jacobian matrix, as reviewed in [22]. Therefore, this paper uses the branch current based DSSE method proposed in [23] for the fault location task. Also, the voltage at the slack node and branch currents are chosen as state variables, and we express these states in a three-phase network as

$$\boldsymbol{x} = [v_{slack,r}^a, v_{slack,x}^a, \cdots, v_{slack,x}^c, i_{1r}^a, i_{1x}^a, \cdots, i_{Lx}^c] \tag{6}$$

where $v_{slack,r}^{\varphi}$ and $v_{slack,x}^{\varphi}$ denote the real and imaginary parts of the $\varphi$ -phase slack node's voltage, and $\varphi \in \{a, b, c\}$; $i_{lr}^{\varphi}$ and $i_{lx}^{\varphi}$ denote the real and imaginary parts of the branch current at branch $l$, $l = 1, \dots, L$, and $L$ is the number of branches. In the following, the phase index $\varphi$ is suppressed for simplicity.

Here, the measurement vector includes the $\mu$PMUs' recorded magnitudes and phase angles of voltages and currents as well as power measurements from pseudo-measurements, and the latter provides the historical or forecasting data with a low-level accuracy of power consumption/production at loads/DGs [21], [22]. We list the measurement functions for voltages, currents, and powers in this estimator as follows:

$$\begin{cases} h_{V_{kr}}(x) = z_j = z_{V_{kr}}, \ k \in \psi_V \\ h_{V_{kx}}(x) = z_j = z_{V_{kx}}, \ k \in \psi_V \end{cases} \tag{7}$$

$$\begin{cases} h_{I_{pr}}(x) = z_j = z_{I_{pr}}, \ p \in \psi_I \\ h_{I_{px}}(x) = z_j = z_{I_{px}}, \ p \in \psi_I \end{cases} \tag{8}$$

$$\begin{cases} h_{P_k}(x) = z_j = z_{P_k}, \ k \in \psi_S \\ h_{Q_k}(x) = z_j = z_{Q_k}, \ k \in \psi_S \end{cases} \tag{9}$$

where $z_j$ denotes measurement $j$ and is expressed as 1) the real and imaginary parts of voltages, $z_{V_{kr}}$ and $z_{V_{kx}}$, 2) the real and imaginary parts of currents, $z_{I_{kr}}$ and $z_{I_{kx}}$, or 3) the real and reactive powers, $z_{P_k}$ and $z_{Q_k}$; $h_{V_{kr}}(x)$, $h_{V_{kx}}(x)$, $h_{I_{pr}}(x)$, $h_{I_{px}}(x)$, $h_{P_k}(x)$, and $h_{Q_k}(x)$ denote the corresponding measurement functions; $\psi_V$ and $\psi_I$ are the sets of nodes and branches with voltage/current measurements from limited $\mu$PMUs installed in the distribution system, and $\psi_S$ is the set of load/DG nodes; $k$ and $p$ are the indices of nodes and branches, respectively. For $k \in \psi_S$, the pseudo-measurements at node $k$ are further converted into equivalent currents in (9) by

$$z_{I_{kr}}^{eq} + jz_{I_{kx}}^{eq} = \left[\frac{z_{P_k}+jz_{Q_k}}{V_k}\right]^* \tag{10}$$

where $z_{I_{kr}}^{eq}$ and $z_{I_{kx}}^{eq}$ are the real and imaginary parts of the equivalent injection current at node $k$; $V_k$ as the voltage phasor at the node is updated during the DSSE procedure, since the $\mu$PMU measurement of $V_k$ is not available at each node; $[\cdot]^*$ denotes the complex conjugate.

By the processing in (10), the Jacobian matrix is independent of $\boldsymbol{x}^t$, *i.e.*, $\boldsymbol{H}(\boldsymbol{x}^t) = \boldsymbol{H}$ [23]. The measurement functions of (7), (8), and (10) and Jacobian elements of $\boldsymbol{H}$ are listed below.

*1) Voltages*

The voltage measurement function of the $\mu$PMU at node $k \in \psi_V$ is expressed as:

$$h_{V_{kr}} + jh_{V_{kx}} = v_{slack} - \sum\nolimits_{p \in \mathfrak{I}_k} (R_p + jX_p) i_p \tag{11}$$

where $\mathfrak{I}_k$ denotes a set of line segments from the slack node to node $k$, and $p \in \mathfrak{I}_k$. $R_p$ and $X_p$ denote the $3 \times 3$ resistance and reactance matrices of branch $p$. Also, the complex variables $i_p = i_{pr} + ji_{px}$ and $v_{slack} = v_{slack,r} + jv_{slack,x}$ are the voltage phasor at the slack node and the current phasor at branch $p$. The Jacobian elements of (11) for $p \in \mathfrak{I}_k$ are constant, expressed as:

$$\frac{\partial h_{V_{kr}}}{\partial v_{slack,r}} = 1 \quad \frac{\partial h_{V_{kr}}}{\partial v_{slack,x}} = 0 \quad \frac{\partial h_{V_{kx}}}{\partial v_{slack,r}} = 0 \quad \frac{\partial h_{V_{kx}}}{\partial v_{slack,x}} = 1$$

$$\frac{\partial h_{V_{kr}}}{\partial i_{pr}} = -R_p \quad \frac{\partial h_{V_{kr}}}{\partial i_{px}} = X_p \quad \frac{\partial h_{V_{kx}}}{\partial i_{pr}} = -X_p \quad \frac{\partial h_{V_{kx}}}{\partial i_{px}} = -R_p$$

*2) Currents*

The current measurement function of the $\mu$PMU at branch $p \in \psi_I$ is shown as

$$h_{I_{pr}} + jh_{I_{px}} = i_{pr} + ji_{px} \tag{12}$$

where $i_{pr}$ and $i_{px}$ denote the real and imaginary parts of the current states, and thus the Jacobian elements are present at:

$$\frac{\partial h_{I_{pr}}}{\partial i_{lr}} = \begin{cases} 1, & when\ p = l \\ 0, & elsewhere \end{cases} \qquad \frac{\partial h_{I_{pr}}}{\partial i_{lx}} = 0$$

$$\frac{\partial h_{I_{px}}}{\partial i_{lr}} = 0 \qquad \frac{\partial h_{I_{px}}}{\partial i_{lx}} = \begin{cases} 1, & when\ p = l \\ 0, & elsewhere \end{cases}$$

where $l$ denotes the branch index, and $l = 1, \dots, L$.

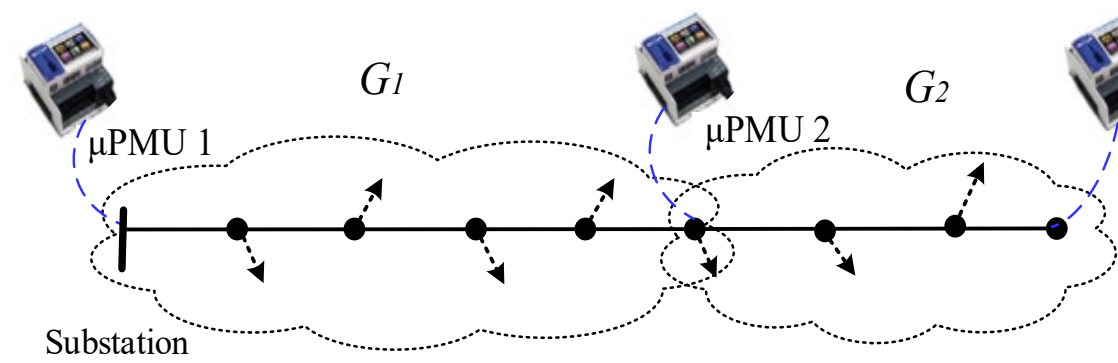

Fig. 1. A sample of radial distribution networks with three $\mu$PMUs. The dotted lines with arrows at nodes denote the laterals (if present), and the $\mu$PMU symbol is from [5].

*3) Power Injections*

The power measurements presenting at node $k \in \psi_S$ are converted into equivalent current injection by (10), and then the current measurement function is expressed as

$$h_{I_{kr}}^{eq} + jh_{I_{kx}}^{eq} = \sum\nolimits_{l \in \Lambda_k'} (i_{lr} + ji_{lx}) - \sum\nolimits_{l \in \Lambda_k} (i_{lr} + ji_{lx}) \tag{13}$$

where $i_{lr}$ and $i_{lx}$ as state variables denote the real and imaginary inflow and outflowing currents at node $k$, and $\Lambda_k'$ and $\Lambda_k$ denote the set of branches with the inflow and outflow currents at node $k$, respectively. The Jacobian elements of (13) are calculated by

$$\frac{\partial h_{I_{kr}}^{eq}}{\partial i_{lr}} = \begin{cases} 1, & when\ l \in \Lambda_k' \\ -1, & when\ l \in \Lambda_k \\ 0, & elsewhere \end{cases} \qquad \frac{\partial h_{I_{kr}}^{eq}}{\partial i_{lx}} = 0$$

$$\frac{\partial h_{I_{kx}}^{eq}}{\partial i_{lr}} = 0 \qquad \frac{\partial h_{I_{kx}}^{eq}}{\partial i_{lx}} = \begin{cases} 1, & when\ l \in \Lambda_k' \\ -1, & when\ l \in \Lambda_k \\ 0, & elsewhere \end{cases}$$

The complete DSSE procedure can be found in [23], and the next section gives the details of the modified DSSE method for faulted line identification.

## III. Graph-Based Faulted Line Identification Method

This section proposes a graph-based fault location method that leverages the above DSSE method to narrow down the searching area and then locate the faulted line.

We consider a distribution network as a graph $G(\mathcal{V}, \mathcal{E})$, where $\mathcal{V}$ and $\mathcal{E}$ denote the sets of vertices (nodes) and the edges (branches), respectively. A $\mu$PMU is installed at the substation, and other $\mu$PMUs are installed at a limited number of nodes along the feeder. Each of these $\mu$PMUs measures the nodal voltage and the currents on the branches connected to that node [23]. We define a subgraph $G_K(\mathcal{V}_K, \mathcal{E}_K)$ as the subset of $G(\mathcal{V}, \mathcal{E})$ that connects two adjacent $\mu$PMUs, $\mu$PMUs $K$ and $K+1$, where $K = 1, \dots, M-1$ and $M \geq 2$. Here, $M$ is the number of $\mu$PMUs installed in the network. Fig.1 shows the schematic diagram of the subgraphs. In the figure, $G_1$ is a subgraph that includes the branches and nodes between $\mu$PMUs 1 and 2, while $G_2$ contains those between $\mu$PMUs 2 and 3.

We briefly describe the proposed fault location method:

1) *Step One:* Using a distributed DSSE algorithm, the searching area for the fault is restricted to the feeder between two adjacent μPMUs, *i.e.*, a certain subgraph.

2) *Step Two*: The location of the fault is further identified as the faulted line.

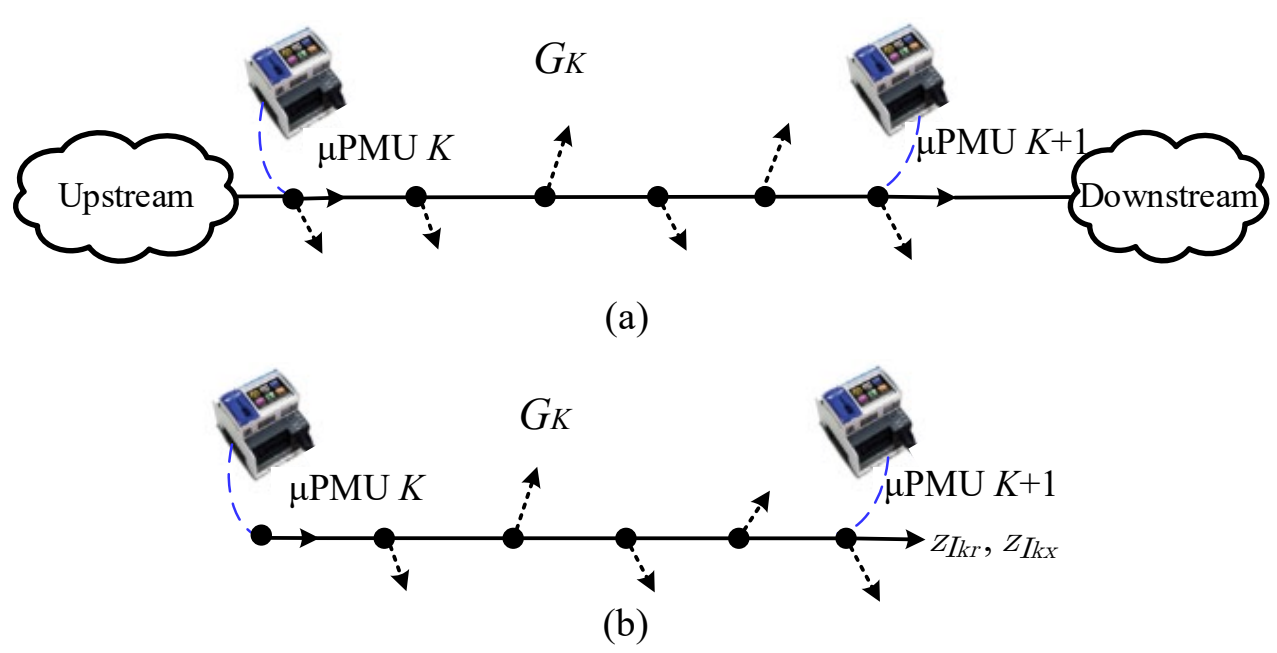

Fig. 2. Subgraph $K$ (a) Embedded in the whole feeder (b) Decoupled with other subgraphs. The reference directions of branch currents measured by two μPMUs are shown.

### A. Step One: Identifying the Faulted Subgraph

This step proposes an efficient DSSE algorithm in $G_K$ to identify the subgraph that contains the faulted line, *i.e.*, the faulted subgraph. By the graph partition and the subsequent network equivalence, the DSSE method leverages the $\mu$PMU and pseudo-measurement data in $G_K$ and runs in parallel for these subgraphs with shorter feeders, *i.e.*, distributed DSSE [24].

*1) Network Equivalence*

In each subgraph, we suppose that the vertex of $G_K$ acts as the root node of this subgraph. The lateral connected to $\mu$PMU $K$+1 is also included in $G_K$, while the lateral at the root node of $G_K$ is included in the last subgraph, *i.e.*, $G_{K-1}$. Fig. 2 shows the schematic diagram of $G_K$ in this design. At node $k \in \mathcal{V}_K$, one type of the following measurements exist and $\mathcal{V}_K = \mathcal{V}_{K1} + \mathcal{V}_{K2} + \mathcal{V}_{K3}$ : 1) Only $\mu$PMU data (*i.e.*, the measurements of $\mu$PMU $K$ at the root node), and let $k \in \mathcal{V}_{K1}$; 2) $\mu$PMU data (*i.e.*, measurements of $\mu$PMU $K$+1) and pseudo-measurements, and $k \in \mathcal{V}_{K2}$; 3) Only pseudo-measurements, and $k \in \mathcal{V}_{K3}$. To reduce the impact of the graph partition on the power flow in the original network shown as Fig. 2(a), we do the equivalent current calculation at node $k \in \mathcal{V}_{K2}$ . Specifically, the real and imaginary parts of the injected current at node $k$ in $G_K$, $z_{I_{kr}}^{sub}$ and $z_{I_{kx}}^{sub}$, are equivalently calculated by

$$z_{I_{kr}}^{sub} + jz_{I_{kx}}^{sub} = (z_{I_{kr}}^{eq} + jz_{I_{kx}}^{eq}) + (z_{I_{kr}} + jz_{I_{kx}}) \tag{14}$$

where $z_{I_{kr}}^{eq}$ and $z_{I_{kx}}^{eq}$ denote the real and imaginary parts of the injection currents of pseudo-measurements obtained by (10), and $z_{I_{kr}}$ and $z_{I_{kx}}$ are the real and imaginary parts of the current to the downstream network measured by $\mu$PMU $K$+1, shown in Fig. 2(b). For simplicity, (14) does not show the measurement noises.

At node $k \in \mathcal{V}_{K1} \cup \mathcal{V}_{K3}$, the measurement functions (10) – (13) hold.

*2) Identification Metric*

We use the WMR in DSSE as the metric to determine the faulted subgraph. In normal operation, assume measurement noises follow Gaussian distribution, WMRs obey a Chi-square distribution with at most $m - n$ degrees of freedom [25, Chapter 5]. With a limited number of $\mu$PMUs installed in distribution systems, the degree of freedom is low and equal to the number of these $\mu$PMUs. Therefore, the values of a WMR in each subgraph fluctuate within a limited range under the impact of measurement noises, when no faults occur.

On the other hand, according to [2], a fault introduces an additional unknown fault current $I_F$ injected to the ground or other phases, while the DSSE equations are built on the precondition $I_F = 0$. If a fault occurs in $G_K$, the presence of the fault violates the state estimation relationship and leads to a high WMR in the faulted subgraph; The DSSE in normal subgraphs have low WMRs even under the impact of measurement noises [19]. Hence, the faulted subgraph is determined by selecting the maximum of WMRs:

$$K^* = \arg \max_K \ J_K \qquad K = 1, \dots, M-1 \tag{15}$$

where $J_K$ denotes the WMR in subgraph $K$ calculated by (2).

Based on the state estimator in (1) – (5), we conclude the procedure for identifying the faulted subgraph below:

i. Considering $G_K(\mathcal{V}_K, \mathcal{E}_K)$, the measurements in each subgraph are collected to form the Jacobian and weight matrices, *i.e.*, $\boldsymbol{H}_K$ and $\boldsymbol{W}_K$.

ii. For $K = 1, \dots, M-1$, the DSSE process in subgraph $K$ is shown in the following steps:

*a. Initialization–forward-backward sweep* [26]: Set the initial voltage at each node as the voltage of the root node $V_{root}$, and calculate the current injections of power measurements by

$$z_{I_{kr}}^{eq} + jz_{I_{kx}}^{eq} = \left[\frac{z_{P_k} + jz_{Q_k}}{V_{root}}\right]^* \tag{16}$$

$$z_{I_{kr}}^{sub} + jz_{I_{kx}}^{sub} = \left[\frac{z_{P_k} + jz_{Q_k}}{V_{root}}\right]^* + (z_{I_{kr}} + jz_{I_{kx}}) \tag{17}$$

where (16) is used for $k \in \mathcal{V}_{K3}$, and (17) holds at $k \in \mathcal{V}_{K2}$; $V_{root}$ comes from the voltage measurement from the $\mu$PMU at the root node.

Then, obtain the initial branch currents $\boldsymbol{x}^0$ by a backward sweep method. Use $\boldsymbol{x}^0$ and $V_{root}$ to calculate initial nodal voltages $V_k$ by a forward sweep method.

*b.* Obtain $\boldsymbol{h}(\boldsymbol{x})$ using (11) – (13), and calculate $\Delta\boldsymbol{x}^t$ and update the new state variables by $\boldsymbol{x}^{t+1} = \boldsymbol{x}^t + \Delta\boldsymbol{x}^t$. Calculate the latest voltages $V_k$ based on the new states by the forward sweep.

*c.* If $\Delta\boldsymbol{x}^t$ is less than a pre-set tolerance or $t$ reaches the maximum iteration number, yield $J_K$ using (2) as the WMR of $G_K$; otherwise, use the latest $V_k$ to calculate injection currents by (10) or (14), then go to step b.

iii. Procure the faulted subgraph $K^*$ using (15).

Finding the faulted subgraph at this stage reduces the computation burden associated with locating the faulted line in subgraphs without faults.

### B. Step Two: Locating the Faulted Line

Once we obtain the faulted subgraph $G_{K^*}$ by Step One, a similar WMR metric based on the DSSE technique is developed to identify the exact line that a fault lies at. Also, we use the following definitions to present Step Two in $G_{K^*}$.

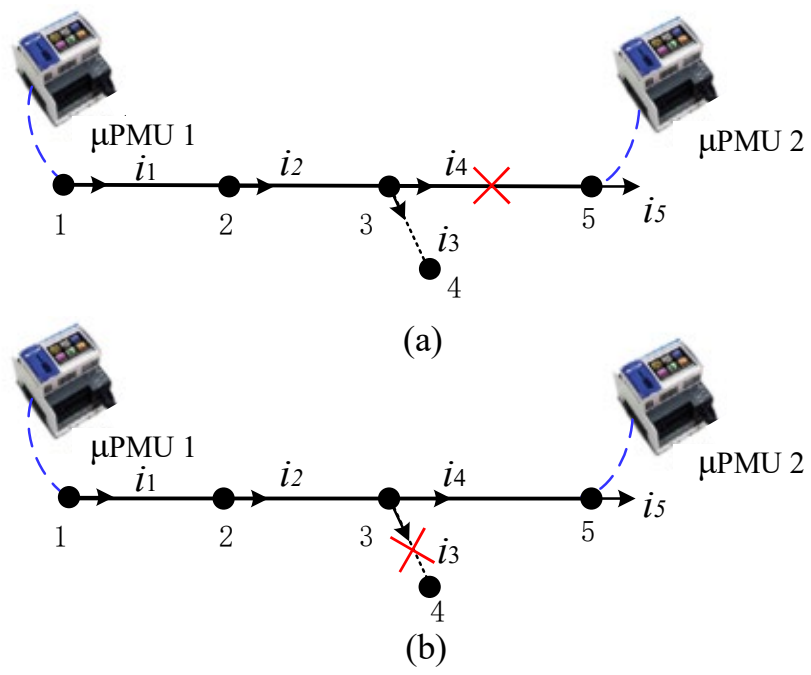

Fig. 3. Sample network of a 5-node subgraph, and a lateral is connected to node 3 shown as a dotted line. (a) A fault occurs at branch 3-5. (b) A fault occurs at branch 3-4.

*Definition 1 (Paths in a subgraph).* A path in a subgraph is a set of interconnected edges that begins with the root node of the subgraph. A path that a fault is located in is a corrupted path.

*Definition 2 (Adjacent Paths and Boundary Edge).* Two paths denoted by $\mathcal{P}_{s-1}$ and $\mathcal{P}_s$, $s = 1,..., S$, are defined as adjacent paths, and if $\mathcal{P}_{s-1} \subseteq \mathcal{P}_s$ and $\mathcal{P}_s = \mathcal{P}_{s-1} \cup \{\varepsilon\}$, where $\varepsilon$ is the boundary edge that connects two vertices $\nu$ and $\mu$, $\nu \in \mathcal{P}_{s-1}$ and $\mu \in \mathcal{P}_s$.

All paths in a faulted subgraph share a starting vertex (root), and different paths are formed by radially expanding the topology of $G_{K^*}$. The paths $\mathcal{P}_s$ in each subgraph are sorted by their depth. The shortest path in the subgraph $K^*$ only includes one edge, while the longest path is the whole subgraph $G_{K^*}$.

In theory, the WMRs in two neighboring paths without fault current injections should be close to each other; the WMR of DSSE in a path is low if there is no fault in the path, while WMR is significantly high once faults occur in the path. Therefore, we convert the fault location problem into a problem of searching for the corrupted path that includes the fault, and this corrupted path is characterized by abnormally high WMR in DSSE. To find the corrupted path, DSSE runs for each path in $G_{K^*}$, and the sending-end branch currents in the corresponding path are chosen as state variables shown in Fig. 3. We apply the DSSE algorithm in Step One for path $s$ and calculate the WMR by

$$WMR_s = [\boldsymbol{z}_s - \boldsymbol{h}_s(\boldsymbol{x})]^T \boldsymbol{W}_s [\boldsymbol{z}_s - \boldsymbol{h}_s(\boldsymbol{x})] \tag{18}$$

where $\boldsymbol{z}_s$ and $\boldsymbol{h}_s(\boldsymbol{x})$ denote the measurement vector and measurement functions for path $s$, and $s = 1,..., S$; $\boldsymbol{W}_s$ denotes the diagonal weight matrix for this path.

According to Definition 2, if a fault occurs at the boundary edge $\varepsilon$, we have

$$WMR_s \gg WMR_{s-1} \tag{19}$$

where $WMR_s$ and $WMR_{s-1}$ are the WMRs in paths $\mathcal{P}_{s-1}$ and $\mathcal{P}_s$ calculated by (2), respectively. To find the faulted boundary edge, set the user-defined identification thresholds to quantize the relationship in (19):

$$\begin{cases} WMR_{s-1} \leq \epsilon \\ WMR_s > \epsilon \end{cases} \tag{20}$$

TABLE I
STATE VARIABLES AND MEASUREMENTS IN A 5-NODE SUBGRAPH

| $s$ | $\mathcal{P}_s$ | State Variables $\boldsymbol{x}$ | Measurements $\boldsymbol{z}_s$ |
|---|---|---|---|
| 1 | 1-2 | $i_1, i_2$ | $z_{V_1}, z_{I_1}, z_{I_2}^{eq}$ |
| 2 | 1-3 | $i_1, i_2, i_3, i_4$ | $z_{V_1}, z_{I_1}, z_{I_2}^{eq}, z_{I_3}^{eq}, z_{I_4}^{eq}$ |
| 3 | 1-5 | $i_1, i_2, i_3, i_4, i_5$ | $z_{V_1}, z_{I_1}, z_{I_2}^{eq}, z_{I_3}^{eq}, z_{I_4}^{eq}, z_{I_5}^{eq}, z_{I_5}, z_{V_5}$ |

where $\epsilon$ denotes the identification threshold for evaluating the abnormally high WMR.

We consider that various fault conditions may occur, and they are unpredictable for system operators. As a result, although a proper identification threshold is beneficial for fault location, the specific value of this threshold is difficult to determine when the fault location, fault impedance, and fault type are unknown. Similar to [2] and [27], the identification threshold $\epsilon$ could be properly selected by using historical or simulation data of different faults to enforce (20). Also, it is efficient to run the efficient DSSE method for verifying the relationship in (20), since the Jacobian matrix $\boldsymbol{H}_s$ for path $s$ is sparse and independent of state variables.

*Illustrative Example*: To clarify the procedure of the proposed method, let us consider a 5-node subgraph shown in Fig. 3, where $l = \{1,2,3,4,5\}$, and a lateral is connected to node 3. There are three paths: 1-2, 1-3, and 1-5 in this subgraph. Path 1-3 is the set of branches from node 1 to node 3, including the lateral 3-4. Table I lists the state variables and measurements used in these paths, and we show $\boldsymbol{H}_s$ for path $s =$ 1, 2, 3, which are marked by three block matrices, respectively. In Fig. 3(a), the fault is located at the boundary edge between paths 2 and 3 by the proposed method, *i.e.*, branch

**Graph-based Faulted Line Identification Algorithm**

**Input:** System model and mesurement data
**While** The presence of a fault is detected, and its location is unknown
  **Step One:** Run the distributed DSSE algorithm for $G_K$ in parellel, and obtain $G_{K^*}$ by (15).
  **Step Two:**
    Let $s = 1$, and obtain $\boldsymbol{z}_s$ and $\boldsymbol{h}_s(\boldsymbol{x})$. Then, calculate $WMR_s$ by (18).
    **If** $WMR_s > \epsilon$
      The faulted line is identified as the first branch in $G_{K^*}$.
    **Else**
      **For** $s = 2{:}\,S$
        **If** $WMR_{s-1} \leq \epsilon$ and $WMR_s > \epsilon$
        The boundary edge between paths $s-1$ and $s$ is located as the faulted line.
        **End if**
      **End for**
    **End if**
**Output:** The faulted line

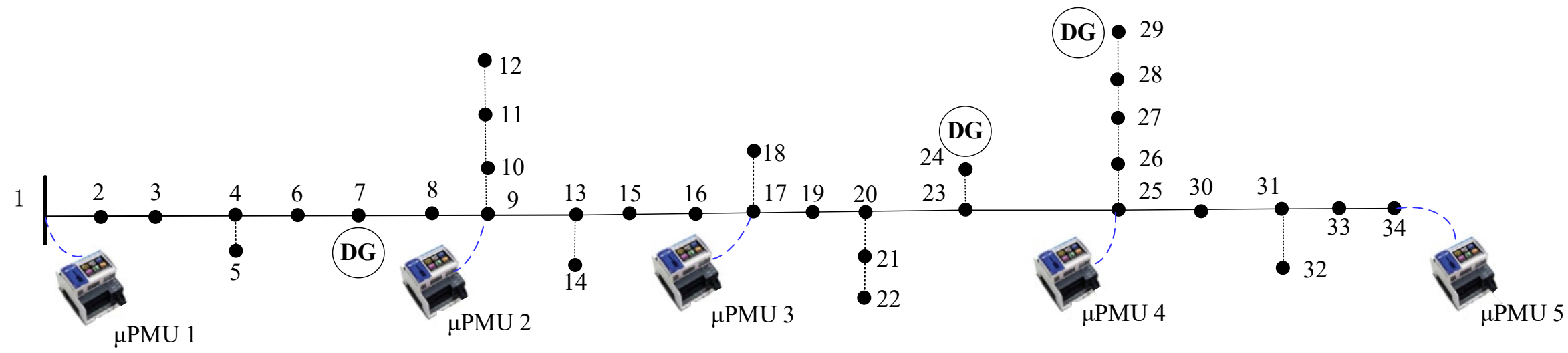

Fig. 4. Single-line diagram for the three-phase 34-node test feeder

3-5; the source of the fault in Fig. 3(b) is located at the boundary edges between paths 1 and 2, *i.e.*, branches 2-3 and 3-4, and there are two boundary edges due to the existence of a lateral. In the case that a fault occurs at a lateral as Fig. 3(b), the fault source could be located at the lateral or the only upstream branch connected to it, even when there is no $\mu$PMU installed at the lateral. Granting complete observability on laterals may not be of economic interest, and therefore, in many practical cases, it is sufficient to identify the faulted laterals [21].

$$
\boldsymbol{H}_s = \begin{array}{c} \\ z_{I_{1r}} \\ z_{I_{1x}} \\ z^{eq}_{I_{2r}} \\ z^{eq}_{I_{2x}} \\ z^{eq}_{I_{3r}} \\ z^{eq}_{I_{3x}} \\ z^{eq}_{I_{4r}} \\ z^{eq}_{I_{4x}} \\ z^{eq}_{I_{5r}} \\ z^{eq}_{I_{5x}} \\ z_{I_{5r}} \\ z_{I_{5x}} \end{array}
\begin{array}{c} i_{1r} \;\; i_{1x} \;\; i_{2r} \;\; i_{2x} \;\; i_{3r} \;\; i_{3x} \;\; i_{4r} \;\; i_{4x} \;\; i_{5r} \;\; i_{5x} \\
\left(\begin{array}{cccccccccc}
1 & & & & & & & & & \\
 & 1 & & & & & & & & \\
1 & & -1 & & & & & & & \\
 & 1 & & -1 & & & & & & \\
 & & 1 & & -1 & & -1 & & & \\
 & & & 1 & & -1 & & -1 & & \\
 & & & & & & 1 & & & \\
 & & & & & & & 1 & & \\
 & & & & & & 1 & & -1 & \\
 & & & & & & & 1 & & -1 \\
 & & & & & & & & 1 & \\
 & & & & & & & & & 1
\end{array}\right) \end{array}
$$

We summarize the proposed algorithm in the pseudo-code. Owing to the hierarchical graph-subgraph-path structure in the proposed method and the adaptation of the advanced DSSE method, the search along the faulted subgraph is more efficient, compared with the ones considering the whole distribution feeder.

### C. *Micro-PMU Placement*

The identification accuracy of the proposed method relies on the number and locations of $\mu$PMUs. As the number of $\mu$PMUs increases, the size of a partitioned graph in the distribution network will be smaller. This would increase the measurement redundancy defined as the ratio of the number of measurements to that of states. The minimum number of $\mu$PMUs to be installed in a distribution feeder is two, similar to the requirement proposed in [21]. Moreover, with more $\mu$PMUs installed, the location performance and computational efficiency of the proposed method can be improved.

TABLE II
NETWORK INFORMATION OF SUBGRAPHS

| subgraph $G_K$ | µPMU $K$ | µPMU $K+1$ | Nodes in $G_K$ |
|---|---|---|---|
| $G_1$ | 1 | 2 | 1-12 |
| $G_2$ | 2 | 3 | 9-18 |
| $G_3$ | 3 | 4 | 17-29 |
| $G_4$ | 4 | 5 | 25-34 |

To guarantee the observability for faulted lines, the following conditions presented in earlier research about meter placement are considered:

1) $\mu$PMU measurements are available at a substation and nodes in the main feeders that have many downstream nodes. Such design is suggested in [24] and [26] to provide improved observability with a limited number of $\mu$PMU measurements.

2) $\mu$PMUs can be installed at the ends of feeders or long laterals to ensure the observability and identify the faulted lateral if necessary [20], [21].

To maximize the location accuracy using a certain number of $\mu$PMUs, an optimal $\mu$PMU placement method presented in [28] can be implemented by considering the probability of various fault types' occurrence. However, the optimal meter placement is a complicated multi-objective optimization problem, involving multiple impact factors, such as the installation cost, estimation accuracy, and faulted line observability, etc. We adopt a meter placement scheme with a low measurement redundancy following the above-mentioned conditions, which illustrates the potential of the proposed method for faulted line identification with limited $\mu$PMU installations.

## IV. CASE STUDY

We test the proposed algorithm on a three-phase unbalanced 24.9kV, IEEE 34-node test feeder [29]. The test system is modified by adding three DGs (two synchronous generators and one PV [20]) into the original systems and simulated in PSCAD, and the proposed location method runs in MATLAB. The detailed models of these DGs can be found in [31]. Fig. 3 shows the $\mu$PMU and DG placement in the system, where five $\mu$PMUs are installed and the nameplate capacity of DGs is 500kVA. The procedure adopted by $\mu$PMUs to obtain voltage or current phasor measurements is described in the appendix. Illustrated as Table II, the propose graph partition in Section III divides the system into four subgraphs, *i.e.*, $G_K \subseteq G$ and $K = 1, 2, 3, 4$.

Assume that measurement noises obey Gaussian distributions. Moreover, the maximum errors of magnitudes and phase angles for $\mu$PMU data are 1% of the true values and 0.01 rads [30], respectively, while the maximum errors for the powers recorded by pseudo-measurements at load/DG nodes are 20% of the true values [25]. Also, smart meters can be installed at DGs for accurately monitoring power outputs, and the maximum errors of these outputs are 3%. By collecting measurements at the DG nodes, distribution system operators (DSOs) do not require the specific DG models. Moreover, DG operators may not share these detailed models and control policies with DSOs due to a lack of agreements between them. However, DSOs can still monitor their power dispatch by the measurement data [7], [31].

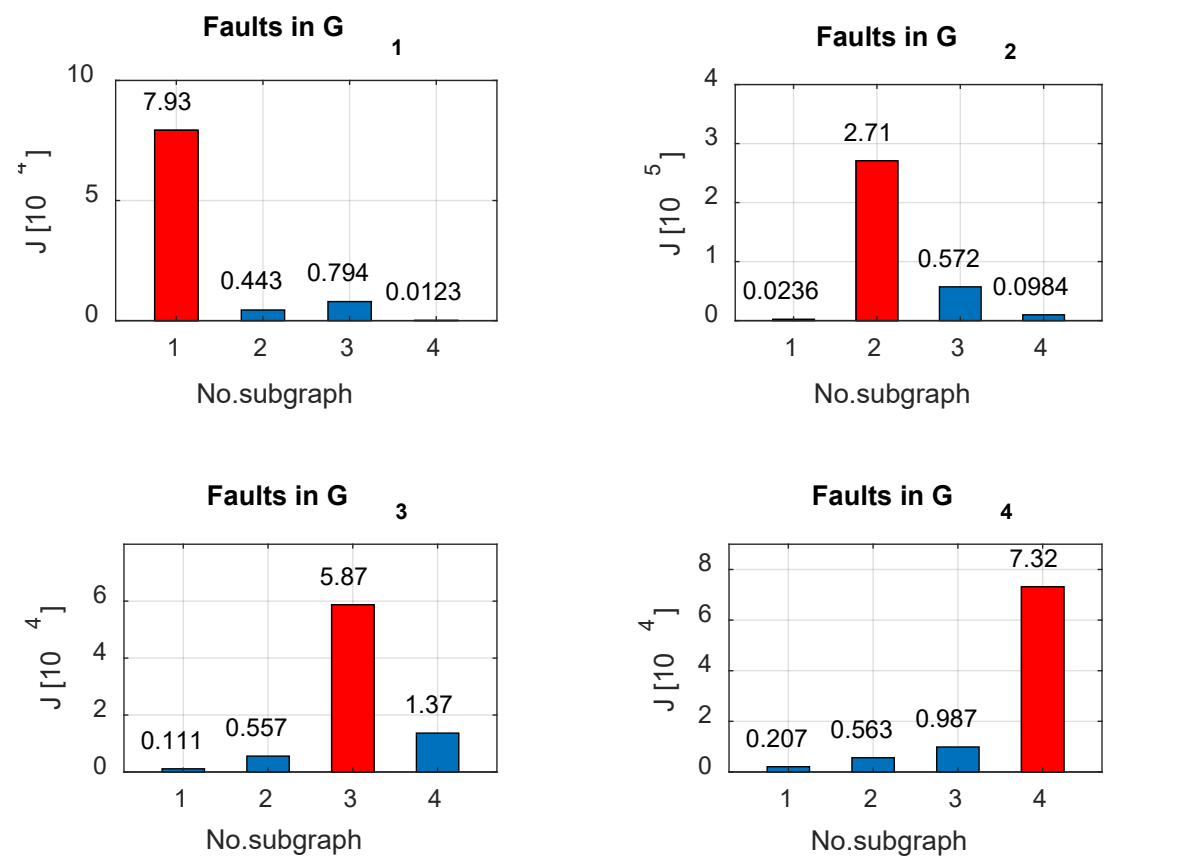

Fig. 5. Identification results in different faulted subgraphs, where we set LG faults on phase A with 100 Ω impedance in $G_K$.

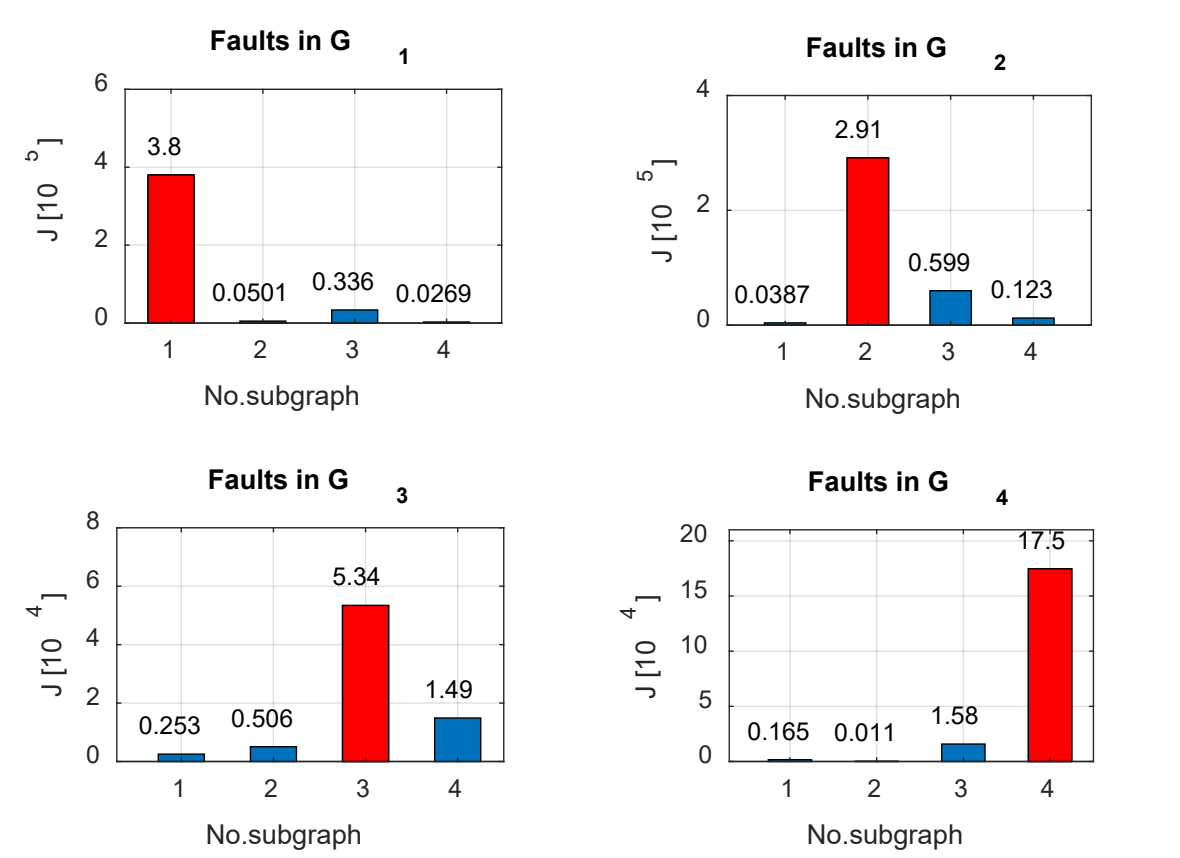

Fig. 6. Identification results in different faulted subgraphs. LL faults on phases B and C with 50 Ω fault impedance.

## A. Faulted-subgraph Identification

This section shows the identification performance of the proposed method in Step One for faulted subgraphs. We test the proposed algorithm with single-phase LG faults, which are set at three branches in each subgraph, *e.g.*, branches 3-4, 7-8, and 10-11 in $G_1$. Moreover, these faults are placed at the beginning ($0.25L_l$), in the middle ($0.5L_l$), and at the end ($0.75L_l$) of the lines, and $L_l$ denotes the corresponding line length. In each fault location, fifty sets of measurements are generated by Monte Carlo simulations. Also, considering nine fault locations for each subgraph, 9×50=450 fault scenarios for two influence factors (fault locations and measurement noises) in each subgraph are tested. In all tests, the subgraphs with the highest values of the identification function $J_K$ correctly point to those faulted subgraphs. The values of $J_K$ across $K$ = 1, 2, 3, 4 for these faults are shown in Fig. 5, where we average the WMRs of each subgraph for conciseness. As discussed in Section III, WMR greatly increases in the faulted subgraph, indicating that the fault is located at that subgraph. For example, when a fault occurs in $G_1$, $J_1$ is abnormally higher than $J_2$, $J_3$, and $J_4$. This leads to an immediate conclusion that the fault is located in $G_1$.

Furthermore, we test the two-phase line-to-line faults with 50 Ω fault impedance in $G_K$, and Fig. 6 depicts $J_K$ in all subgraphs. We observe that the maximum of $J_K$ correctly indicates the location of the faulted subgraph. Also, the identification performance for the faulted subgraph is not influenced by fault types and fault impedances.

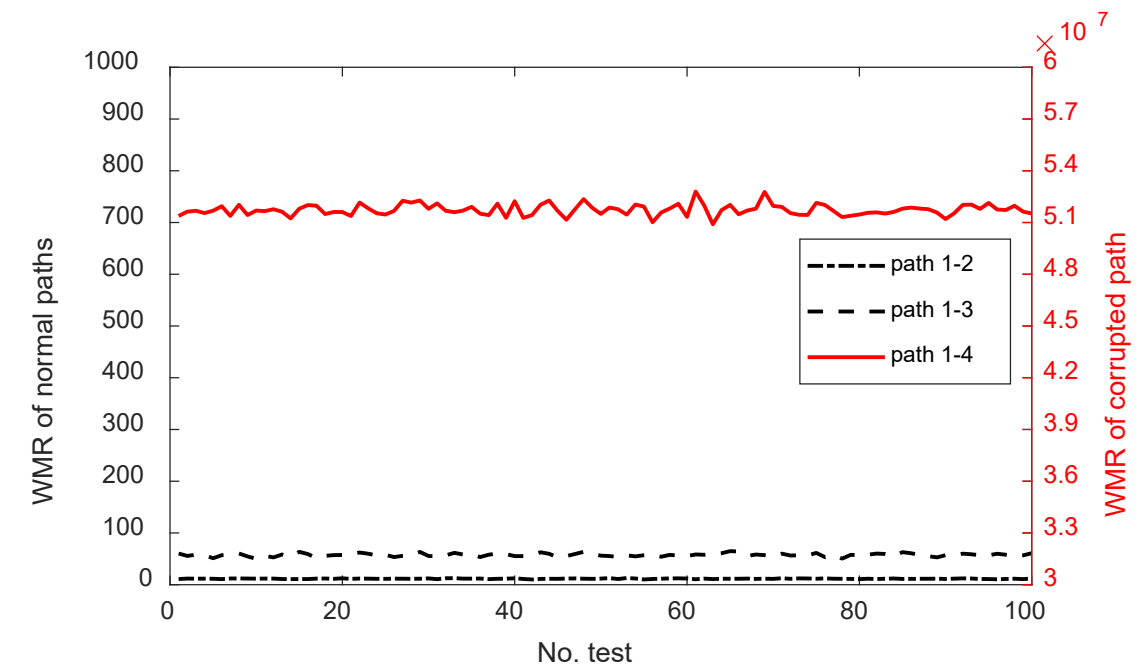

Fig. 7. Location results in $G_1$, when LL faults with 10Ω fault impedance occur at branch 3-4. The secondary y-axis shows the WMRs at the corrupted path in 100 Monte Carlo simulations.

TABLE III
PERFORMANCE WITH DIFFERENT FAULT TYPES (50Ω IMPEDANCE)

| Fault Type | $\alpha$ | $\beta$ | $1-\alpha-\beta$ | Max Error |
|---|---|---|---|---|
| LG | 94.50% | 4.67% | 0.83% | 2 branches |
| LL | 95.75% | 4.25% | 0% | 1 branch |
| LLG | 96.83% | 3.17% | 0% | 1 branch |
| LLL | 95.67% | 4.33% | 0% | 1 branch |

## B. Faulted-line Location

We test various fault scenarios to evaluate the location performance of the proposed method. Fig.7 shows the WMRs for different paths in the faulted subgraph for LL faults on phases B and C on branch 3-4 in $G_1$, where we run 100 Monte Carlo simulations for random combinations of measurement noises. In this figure, the WMRs of the normal paths are much lower than those for the corrupted paths. Also, with the radial expansion of paths, the WMRs of the corrupted path that the boundary edge 3-4 lies in have high values. Consequently, branch 3-4 is identified as the faulted line.

TABLE IV
PERFORMANCE WITH DIFFERENT FAULT IMPEDANCES (LL FAULTS)

| Fault Impedance | $\alpha$ | $\beta$ | Max Error |
|---|---|---|---|
| 0 Ω | 100% | 0% | 0 branch |
| 10 Ω | 94.67% | 5.33% | 1 branch |
| 50 Ω | 94.83% | 5.17% | 1 branch |
| 100 Ω | 100% | 0% | 0 branch |
| 200 Ω | 95.08% | 4.92% | 1 barnch |

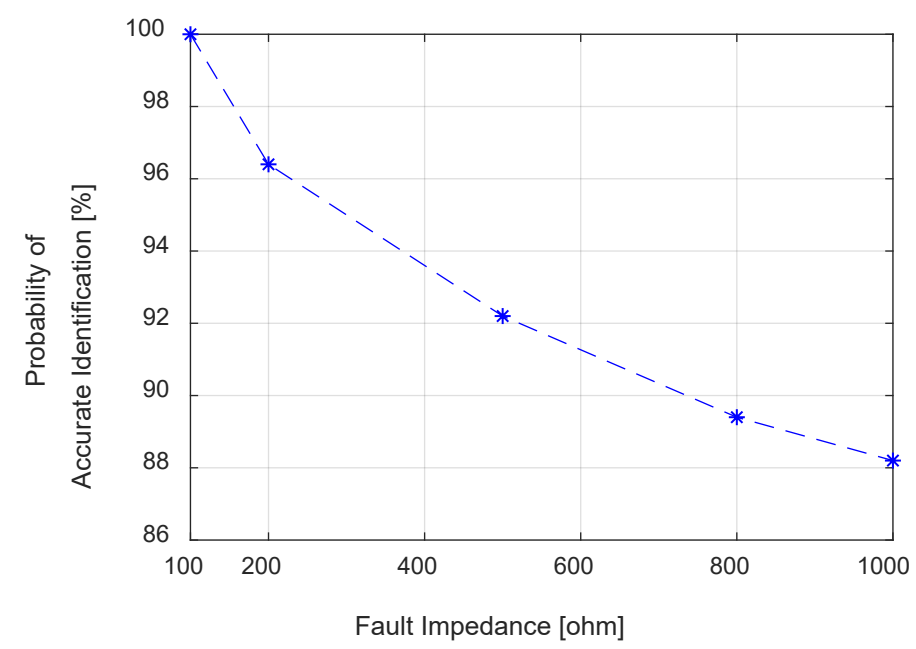

Fig. 8. Performance for high-impedance LG faults at branch 3-4

We further evaluate the location accuracy of the proposed method by calculating the probabilities of two types of test results: 1) the faulted branch is correctly located and 2) an immediate neighboring branch of the faulted branch is determined as a faulted one [21]:

$$\alpha = N_0/N_t \tag{21}$$

$$\beta = N_1/N_t \tag{22}$$

where $N_0$ and $N_1$ denote the number of the tests in these two cases, respectively, and $N_t$ is the total number of the tests; also, $1-\alpha-\beta$ is used to calculate the probability of other results, *i.e.*, other branches are determined as a faulted line.

We calculate these accuracy indices $\alpha$ and $\beta$ in scenarios with various fault types and fault impedances, where $N_t = 1200$ is set to obtain statistical results in each scenario, and here the identification threshold $\epsilon = 500$.

*1) Fault Type*

The impacts of various fault types on the location accuracy of the proposed algorithm are investigated. Four fault types denoted as LG, LL, LLG, and LLL, are tested. We list the location results of these fault types in Table III. It is shown that the proposed method enables correct faulted line location with various fault types and reaches 94% and higher accuracy.

*2) Fault Impedance*

We test the impacts of fault impedances on the accuracy of the proposed algorithm. We set different fault impedances at each branch of $G_2$, and Table IV shows the accuracies of this method to locate faults with these impedances. Especially, the proposed method enables accurate locations of bolted faults, owing to the existence of the fault injection currents with high magnitudes. The results in Table IV show that the proposed method enables correct fault-line location with multiple fault impedances.

TABLE V
PERFORMANCE WITH HIGHER MEASUREMENT ERRORS (LG, 50 Ω)

| Max Error of μPMU Data | Max Error of Pseudo-meas. | $\alpha$ | Max Error |
|---|---|---|---|
| 2%, 0.02 rads | 10% | 94.67% | 1 branch |
| | 30% | 94.50% | 1 branch |
| | 50% | 94.42% | 1 branch |

TABLE VI
IMPACT OF IDENTIFICATION THRESHOLDS

| Threshold $\epsilon$ | $\alpha$ | $\beta$ | $1-\alpha-\beta$ | Max Error |
|---|---|---|---|---|
| 100 | 88.69% | 10.35% | 0.96% | 2 branches |
| 500 | 95.92% | 3.91% | 0.17% | 2 branches |
| 1000 | 97.08% | 2.92% | 0% | 1 branch |
| 2000 | 97.25% | 2.75% | 0% | 1 branch |

The proposed method is tested with high-impedance LG faults (100, 200, 500, 800, and 1000 Ω) at branch 3-4 to show the sensitivity towards magnitudes of fault currents. Fig.8 shows the probabilities of the correct location of the faulted branch, where Monte Carlo simulations with 400 samples of measurements are used. The location probabilities are higher than 88% under these various current injections. The reason is that the measurement errors of voltages and currents are proportional to the measurement values, while the measurement weights are inversely proportional to them. While smaller fault current injections occur, as the weights of measurements are higher in this case, the WMR will be high. We conclude that the proposed method works effectively when the fault impedance is not higher than 1000 Ω in the test system. Once the fault impedance exceeds about 2000 Ω, the proposed approach may not observe the small fault injection at long branches in the 34-node system.

## C. Robustness and Sensitivity Analysis

We investigate the robustness and sensitivity of the proposed method against various measurement noises and identification thresholds.

*1) Measurement Errors*

We conduct robustness analysis concerning higher measurement noises. We set the measurement noises of μPMUs as 2% in magnitudes and 0.02 rads in phase angles, while considering the maximum errors of pseudo-measurements as 10%, 30%, and 50%. Table V lists the accuracy of the proposed algorithm with these measurement noises. As shown, even with high pseudo-measurement errors up to 50%, either the correct line or its immediately neighboring line is identified. It implies that such high-level noises do not degrade the location performance since DSSE takes the weights of measurement noises into full account. Also, the location performance of this algorithm is robust against measurement errors.

*2) Identification Threshold*

We test the location performance of the proposed method with various identification thresholds $\epsilon$. Fig. 7 shows that the WMRs in the corrupted paths are much higher than those for the normal paths. Further, different thresholds are set in the cases of Section IV-B, and the location accuracy with these thresholds is calculated and listed in Table VI. We conclude that the identification threshold could be properly selected to maintain a desirable identification sensitivity.

TABLE VII
LOCATION ACCURACY WITH UNCERTAINTY IN LINE PARAMETERS

| Maximum Errors of Line Parameters | $\alpha$ | $\beta$ | Max Error |
|---|---|---|---|
| 2% | 100% | 0% | 0 branch |
| 5% | 97.5% | 2.5% | 1 branch |
| 10% | 95.5% | 3.25% | 2 branches |

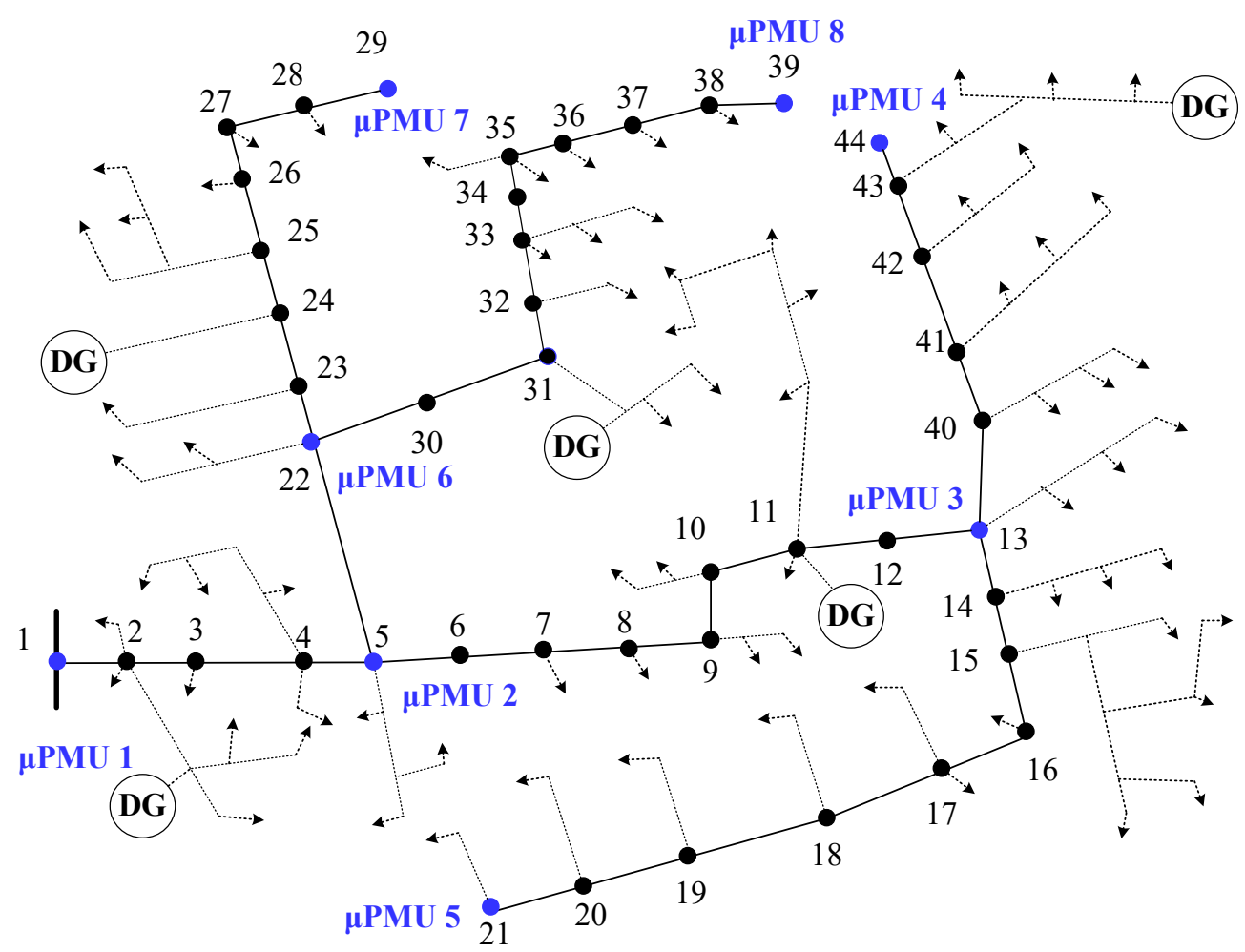

Fig. 9. Diagram of the modified 123-node distribution system

## D. Impact of Line Parameters

Line parameters in distribution systems are subject to changes with environmental conditions. Considering this uncertainty, the range of line parameters is generally set within ±5% of their nominal values [32]. Therefore, we consider the variation in line parameters to evaluate the accuracy of the proposed method.

We use Monte Carlo simulations to generate 400 test scenarios, where imprecise line parameters are assumed to obey Gaussian distribution with various maximum deviations and zero means. Table VII lists the location accuracy of the proposed algorithm, and the maximum errors are 2%, 5%, and 10% of true values of line parameters. Also, LG faults with 100 Ω are set on different branches in $G_2$. With 5% deviation in the line parameters, either the correct faulted line or its immediate neighboring branch is identified.

We conclude that inaccurate line parameters degrade the location accuracy of the proposed method, and hence line calibration in power systems is necessary periodically.

## E. Performance in a Larger-scale System

To show the scalability of the proposed graph-based algorithm, we test the proposed method on the modified IEEE 123-node distribution system, where the nodes at the main feeders are renumbered by [21]. The details of the 123-node system can be found in [29]. Fig.9 depicts the DG installations and meter placement, and $\mu$PMUs are placed according to the conditions in Section III-C. The distribution network graph is divided into six subgraphs as shown in Table VIII. As the faults on branch 5-22 are directly observable by two-end $\mu$PMUs, and thus this branch is not included in the subgraphs in Table VIII.

TABLE VIII
SUBGRAPH INFORMATION IN 123-NODE SYSTEM

| subgraph $G_K$ | Nodes in $G_K$ |
|---|---|
| $G_1$ | 1-5 between µPMUs 1 and 2 |
| $G_2$ | 5-13 between µPMUs 2 and 3 |
| $G_3$ | 13-21 between µPMUs 3 and 4 |
| $G_4$ | 13-44 between µPMUs 3 and 5 |
| $G_5$ | 22-29 between µPMUs 6 and 7 |
| $G_6$ | 22-39 between µPMUs 6 and 8 |

TABLE IX
LOCATION ACCURACY IN THE 123-NODE SYSTEM

| Fault Type | $\alpha$ | $\beta$ | Max Error |
|---|---|---|---|
| LG | 93.67% | 4.67% | 2 branches |
| LL | 95.67% | 4.33% | 1 branch |
| LLG | 96.33% | 3.67% | 1 branch |
| LLL | 96.0% | 4.0% | 1 branch |

Table IX lists the accuracy of fault location for various fault types in subgraphs $G_1$, $G_3$ , and $G_5$, where the fault impedance is set as 50 Ω. Set $N_t = 300$ for each fault type to calculate $\alpha$ and $\beta$ by (21) and (22), and set $\epsilon = 2000$. Table IX shows that the proposed method can correctly identify the faulted lines with accuracy higher than 93%. We conclude that the proposed algorithm uses the distributed DSSE technique to work in this larger-scale distribution system.

## F. Computational Efficiency

Numerical experiments for different faults are performed to demonstrate the computational efficiency of the proposed algorithm. We run this method on a PC with 2.6 GHz i5, and 8GB RAM using MATLAB 2017b.

Table X lists the average CPU time of the proposed method, including two steps, for faulted line identification in these test systems. It shows that once the measurement data are collected, this algorithm locates the faults within 15 milliseconds in the 34-node distribution system. Compared to the traversal search strategy for a whole feeder in [19] and [21], the proposed method runs in parallel for feeders with a reduced size, which improves the computational efficiency for application to the larger-scale networks. Hence, our location method has a low computational cost, which is also verified in the 123-node test system. It should be noted that owing to the increase in the number of nodes in subgraphs of this larger-scale system, the proposed method takes a longer CPU time, *i.e.*, about 20 ms, for faulted line identification. With more $\mu$PMUs installed, the number of nodes in a subgraph decrease, and the computational efficiency of this method can be further improved.

# V. CONCLUSION AND OUTLOOK

This paper proposes a graph-based faulted line identification algorithm using $\mu$PMU data in distribution systems. We present a distributed DSSE algorithm to identify the faulted subgraph efficiently, and this method significantly reduces the searching scale and speeds up the subsequent fault location procedure. Further, we conveniently determine a faulted line by applying a hierarchical graph-subgraph-path structure to the DSSE method. In the case of inadequate $\mu$PMU and DGs installed at the distribution level, the proposed method enables accurate faulted-line location. Extensive simulations verify the accuracy and efficiency of this method under various fault scenarios.

TABLE X
CPU TIME IN TWO TEST SYSTEMS

| #System | Faulted Subgraph | Average CPU Time [ms] |
|---|---|---|
| 34-node System | $G_1$ | 14.28 |
| | $G_2$ | 13.50 |
| | $G_3$ | 14.72 |
| | $G_4$ | 9.98 |
| 123-node System | $G_1$ | 21.37 |
| | $G_3$ | 22.69 |
| | $G_5$ | 17.92 |

## APPENDIX

The proposed method leverages post-fault phasors from $\mu$PMUs. Fig.10 shows the diagram of a waveform for fault currents and the latency of the proposed method for faulted line identification. According to [19], considering the duration of transients, *i.e.*, $T_1$, there is a short latency between 0 and 20 ms before measuring the steady-state phasors by $\mu$PMUs. Later, to obtain accurate post-fault synchrophasors, the discrete Fourier transform (DFT) method is used to process a dataset of raw-sampled waveforms [2]. The time window $T_2$ for $\mu$PMUs to get post-fault phasors is about two to three periods derived by a fundamental frequency, *e.g.*, 30 ms. Moreover, according to IEEE Standard C37.118, the shortest length of the observation window can reach up to 17 ms [30].

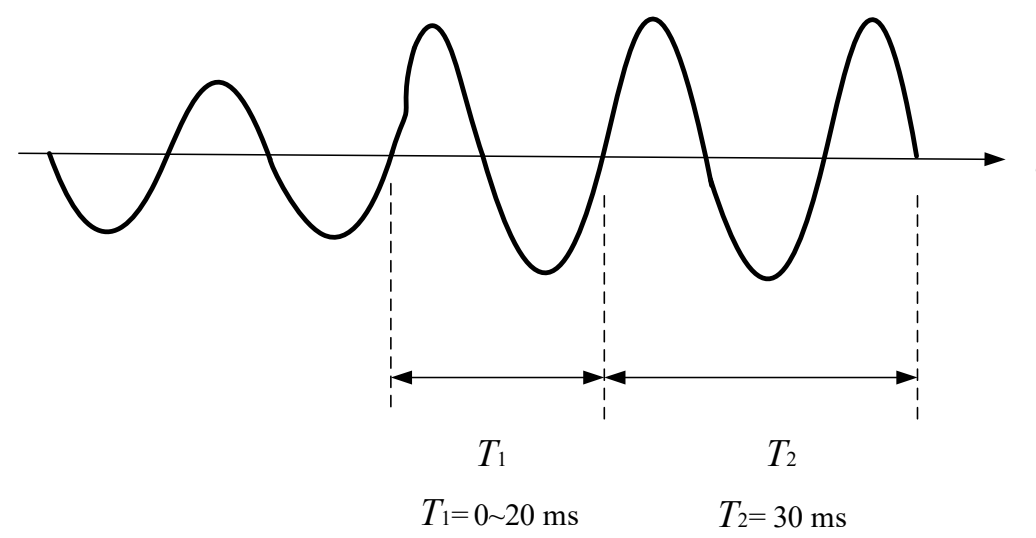

Fig. 10. The diagram of a waveform of fault current and the latency of the proposed method in identifying a fault [19].